\documentclass[aps,prc,preprint,showpacs,nofootinbib]{revtex4}

\usepackage{epsfig}
\usepackage{graphicx,amsmath}
\newcommand\ba{\begin{eqnarray}}
\newcommand\ea{\end{eqnarray}}

\newcommand{\be}{\begin{equation}}
\newcommand{\ee}{\end{equation}}

\newcommand{\bas}{\begin{eqnarray*}}
\newcommand{\eas}{\end{eqnarray*}}
\begin{document}
\title{\bf \large Exotic low mass narrow baryons extracted from charge exchange reactions}

\author{
B. Tatischeff$^{1,2}$\thanks{e-mail : tati@ipno.in2p3.fr}
 and E. Tomasi-Gustafsson$^{1,2,3}$\thanks{e-mail :etomasi@cea.fr}\\
$^{1}$CNRS/IN2P3, Institut de Physique Nucl\'eaire, UMR 8608, Orsay, F-91405\\
$^{2}$Univ. Paris-Sud, Orsay, F-91405, France\\
$^{3}$ IRFU/SPhN, CEA/Saclay, 91191 Gif-sur-Yvette Cedex, France\\
}
\pacs{}

\vspace*{1cm}
\begin{abstract}
This paper aims to give further evidence for the existence of low mass exotic baryons. Narrow structures in baryonic missing mass or baryonic invariant mass were observed during the last twelve years. Since their evidence is still under debate, various data, measured with incident hadrons, by different collaborations, are reanalyzed to bring evidence on  these narrow exotic baryonic resonances excited in charge-exchange reactions.  These structures are clearly exotic as there is no room for them in the $qqq$ configurations \cite{caps}: their width is smaller than the widths of "classical" baryonic resonances \cite{pdg}, moreover some of the masses lie below the pion threshold mass. 
\end{abstract}
\maketitle
\section{Introduction}
Already observed since several years, the experimental observation of narrow low-mass baryonic structures is not yet accepted as an evidence. This paper summarizes briefly the experiments where these structures were observed, and reanalyses other data, in order to add information on these narrow baryonic exotic resonances.

Although obtained with different physical motivations, therefore with statistical precision lower than the precision preferable for the present study, these results  add information on these structures.
We reanalyze here cross-sections data involving only hadrons. Indeed, the narrow structures were associated with multi-quark clusters, which may explain that they are hardly observable in reactions involving incident leptons. Concerning pion charge-exchange reactions, the statistics is too poor for the present study. 

Therefore, in the present work, several spectra on baryon charge-exchange reactions are reanalyzed, and plotted as a function of the missing masses even though they were originally published as a function of the energy loss.
\section{Brief recall of previous results}
\subsection{The SPES3 (Saturne) data}
Previous experiments, performed at SPES3 (Saturne), which benefit of good resolution and high statistics, exhibit narrow structures at different 
hadronic masses. Only results concerning baryons are discussed here.  In particular, two reactions: 
\be
p+p\to p+p+X   
\label{eq:eq1}
\ee
and
\be
p+p\to p+ \pi^++X  
\label{eq:eq2}
\ee
were studied \cite{bor1,bor2}.  Structures were observed in the missing mass M$_{X}$ of reaction (\ref{eq:eq2}), in the invariant mass M$_{pX}$ of reaction (\ref{eq:eq1}), and  in the invariant masses M$_{p\pi^{+}}$ and M$_{\pi^{+}X}$ of reaction (\ref{eq:eq2}). 
\begin{center}
\begin{figure}[!ht]
\caption  {Missing mass differential cross-sections for pp$\rightarrow$ p${\pi^+}$X 
reaction for the three lowest angles at T$_p$=1520 and 1805~MeV, 
 selected for missing masses larger than 960 MeV.  Data measured at SPES3 (Saturne)      \protect\cite{bor1}. They have been
binned into 5.6~MeV mass intervals, shifted and rescaled in order to present the six angles inside the same figure. Vertical lines 
indicate the mean position of the structures ($\alpha$) M=1004~MeV, ($\beta$) 
M=1044~MeV, and ($\gamma$) M=1094~MeV.}
\label{fig1}
\end{figure}
\end{center}
\vspace*{-0.8cm}
Narrow structures were observed in different conditions (reaction, incident energy, spectrometer angle, or observable) at the same masses (within $\pm$3~MeV) which is, in the opinion of the authors, a strong indication of their existence.

A single figure (Fig. \ref{fig1}) is reproduced here, since it corresponds to the first observation of these exotic baryons \cite{bor1}, and since the corresponding masses, narrow and lying below the pion threshold mass, prove directly the exotic nature of these states.
We observe three structures very well defined statistically. Several other narrow structure exotic masses were extracted from different studies done at SPES3 (Saturne). They were published in \cite{bor2,bor3,bor4},  showing data in the missing mass range: 1.0$\le$M$\le$1.46~GeV, 1.47$\le$M$\le$1.68~GeV, and 1.72$\le$M$\le$1.79~GeV, respectively. 

The values of the observed narrow structures masses are: 1004($\alpha$), 1044($\beta$), 1094($\gamma$), 1136($\delta$), 1173($\epsilon$), 1210($\lambda$),
1249($\eta$), 1277($\phi$), 1339($\nu$), and 1384~MeV. In this work, the greek letters within parenthesis, identify the structures shown in the figures and the vertical arrows show the expected position for these narrow structures on the basis of previous observations. Some narrow structures in the mass range M$_{N}\le $M$\le$1~GeV, were also tentatively extracted \cite{doubna2}. 
 
Other signatures of narrow baryonic structures, were observed either in dedicated experiments or  extracted from cross sections obtained and published by different authors. They are quoted in \cite{bor2} and will not be recalled here.
\section{The p(d,d')X reaction}
This reaction, like the next one p($\alpha,\alpha'$)X,  was studied in order to get information on the Roper resonance excitation.
Less precise cross-sections for the p(d,d')X reaction were measured at Saturne (SPES4) \cite{banaigs1,baldini} using incident deuteron beam with momentum p$_{d}$=2.95~GeV/c, at  five laboratory spectrometer angles, from $\theta=6.59^{\circ}$ up to $11.95^{\circ}$. The model predictions  \cite{baldini,hirenzaki} are rather far from the data. 

The reanalysis of the two smallest angles: $\theta=6.59^{\circ}$ and $8.05^{\circ}$, respectively in inserts (a) and (b), is shown in Fig. \ref{fig2}. 
The broad curve is plotted to drive the eyes through the small triangle points which reproduce the result of the theoretical calculation \cite{hirenzaki} including the contribution from $\Delta$ and Roper excitations. We believe that the significant gap between this calculation and data can be filled by the exotic baryonic structures. For this aim we take the fitting function as a sum of gaussians centered at fixed values quoted above, with a constant width. The only free parameters are the relative normalization. This allows to get a good fit to the data. The expected structure around M=1320~MeV is absent; it will also be the case for other fits shown later on.
\begin{center}
\begin{figure}[!ht]
\caption {(Color online). Cross-section of the p(d,d)X reaction at p$_{d}$=2.95~GeV/c, \protect\cite{baldini} measured at Saturne (SPES4 beam line). Inserts (a) and (b) show respectively the results for $\theta=6.59^{\circ}$ and $8.05^{\circ}$. The theoretical calculation \protect\cite{hirenzaki} is shown by the small triangles and the thin curve to drive the eyes.}
\label{fig2}
\end{figure}
\end{center}
\section{The p($\alpha,\alpha'$)X reaction} 
Large statistics spectra of the p($\alpha,\alpha'$)X reaction were obtained several years ago at SPES4 (Saturne) in order to study the radial excitation of the nucleon in the P$_{11}$(1440~MeV) Roper resonance. The experiment was performed with T$_{\alpha}$=4.2~GeV. 
 A spectrum measured at
T$_{\alpha}$=4.2~GeV, $\theta=0.8^{\circ}$ \cite{morsch1} was obtained in the
baryonic missing mass range: 1030$\le$M$\le$1490~MeV. Fig. \ref{fig3} shows these data for the cross-section.  A  first large peak around $\omega\approx$~240~MeV was
associated with the projectile excitation, and a second large peak around $\omega\approx$~510~MeV was associated with the target excitation \cite{morsch1}. Here $\omega$ is the energy difference between the incident and the detected $\alpha$ particles.

 Above these large peaks lie narrow peaks, clearly observed, defined by a large number of standard deviations \cite{doubna2004} (see Fig. \ref{fig3}) .

 A second spectrum with large statistics was measured at T$_{\alpha}$=4.2~GeV, $\theta=2^{\circ}$ \cite{morsch2} which extended up to M=1588~MeV. Fig. \ref{fig4} shows this data. In both figures, the empty circles, which correspond to the scale, are the published number of events versus the missing mass. The full circles and full squares show the same data with magnified scale.

The narrow structure masses agree fairly well with the masses of narrow structures for the 
pp$\to$p$\pi^{+}$ and pp$\to$ppX reactions studied at SPES3 (Saturne) \cite{bor2}.
Table 1 gives the quantitative information concerning the masses extracted from the previous figures, and the comparison with the masses from SPES3 cross-sections
\cite{bor2}. Nearly all peaks are visible in both experiments. At
$\theta=0.8^{\circ}$, the incident beam enters in the SPES4 spectrometer,
preventing a possible confirmation of the lower mass structure at 
M=1004~MeV.  At $\theta=2^{\circ}$, Fig. \ref{fig4}, several peaks are observed above M=1470~MeV. 
Fig. \ref{fig5} shows an enhanced plot of this missing mass range ($\omega\ge$800~MeV). The same situation, with many structures, is observed in the SPES3 data \cite{bor3,bor4}, and the masses observed in both reactions are consistent. The peak at M=1394~MeV, observed in the SPES4 experiment, was not seen in the SPES3 data, since such missing mass range lies between two incident proton energies.  The statistical errors of the  p($\alpha,\alpha'$)X cross-sections, could not be larger than a factor of two, calculated from the statistical errors \cite{morsch3}. The error bars are therefore multiplied by a factor of two. 
\begin{center}
\begin{figure}[!ht]
\caption{Cross-section of the p($\alpha,\alpha'$)X reaction at T$_{\alpha}$=4.2~GeV, $\theta=0.8^{\circ}$ \protect\cite{morsch1}. Full points show magnified data.}
\label{fig3}
\end{figure}
\end{center}
\begin{center}
\begin{figure}[!ht]
\caption{Cross-section of the p($\alpha,\alpha'$)X reaction at T$_{\alpha}$=4.2~GeV, $\theta=2^{\circ}$ \protect\cite{morsch2}. Full points show magnified data.}
\label{fig4}
\end{figure}
\end{center}
We observe a fair agreement between the masses obtained
using data originally collected in different experiments, studied with different aims, carried out by different physicists in different reactions with
different probes and different experimental equipements.

This agreement is illustrated in Fig. \ref{fig6} where the points are well aligned along straight lines which correspond to the same masses. 
\section{New reanalysis of various different reactions}
A large number of charge-exchange reactions were studied in different laboratories, mainly in Saturne (SPES4 beam line) with the aim to study isospin-spin excitations. 
The corresponding missing mass data range from nucleon up to $\Delta$(1232). Therefore these data are quite convenient for the present study. An important part of these data is  presented in the following paragraphs and figures. Some preliminary reanalysis were previously shown \cite{varenna}. In the following spectra, we add to the broad $\Delta$(1232) bump, several narrow gaussians describing the structures observed previously, at {\it the masses previously fixed } by the analysis of the  
cross sections of the $p+p\to p+ \pi^{+}+X$ and   $p+p\to p+p+X$ reactions studied at SPES3(Saturne).
\begin{center}
\begin{figure}[!ht]
\caption{Same caption as in Fig. \protect\ref{fig4}}
\label{fig5}
\end{figure}
\end{center}
\begin{center}
\begin{figure}[!ht]
\caption{(Color online.) Comparison between masses of narrow baryons extracted from
SPES3 \cite{bor1,bor2,bor3}and SPES4 \cite{morsch1,morsch2} data. Inserts (a) and (b) correspond respectively to
$\theta=0.8^{\circ}$ and $\theta=2^{\circ}$. }
\label{fig6}
\end{figure}
\end{center}
\section{The (p,n) reactions}
\subsection{The pp$\to$n$\Delta^{++}$ reaction.}
The cross-sections of the pp$\to$n$\Delta^{++}$ reaction were measured at different incident energies in different laboratories, and analyzed phenomenologically within the one-pion exchange model \cite{dimitriev}.
The binning and statistics of these old data are not very precise, therefore only one spectrum is shown. Fig. \ref{fig7} shows the data at $T_{p}$=0.97~GeV \cite{bugg}, the result of the calculation performed \cite{dimitriev} (with $\Lambda_{\pi}$=0.63~GeV),  and a few gaussians introduced as previously. Adding such gaussians appears to be essential for an accurately the data. These gaussians have all the same width ($\sigma$=12~MeV), and their positions are fixed at the masses where narrow exotic baryons were previously observed. The addition to $\Delta$ peak of gaussians describing the baryonic structures found in previous works, allows to get a very good  description of the experimental data. The calculation performed in  \cite{dimitriev,bugg} of the $\Delta^{++}$ peak alone, is unable to fully describe the data. On the one hand, the statistical significance between data and broad PDG $\Delta$ resonance is $\chi^{2}_{D}$=4.7 on the other hand, the statistical significance between data and total spectra including the narrow resonances is $\chi^{2}_{NR}$=2.1 (see Table 2).
\begin{center}
\begin{figure}[!ht]
\caption{(Color online.) Cross-section of the  pp$\to~n\Delta^{++}$ reaction at $T_{p}$=0.97~GeV $\theta=0^{\circ}$ \protect\cite{dimitriev}}. 
\label{fig7}
\end{figure}
\end{center}
More recent data of the $^{1}$H(p,n)X reaction exist at T$_{p}$=790~MeV,
$\theta_{n}$=0$^{\circ}$, 7.5$^{\circ}$, and 15$^{\circ}$. The corresponding differential cross-sections $d^{2}\sigma/dE_{n}d\Omega_{n}$
were measured at LAMPF \cite{prout}. These data at the two smallest angles are reanalyzed  in Fig. \ref{fig8}.  The full curves fitting the small empty triangles, describe the theoretical calculations \cite{mosbacher}
performed with use of an effective projectile-target-nucleon interaction. These curves are reproduced here just as the data; they describe the broad $\Delta$ peak. A few points corresponding to the largest energy loss ($\omega$) values are omitted, since they lie over the Jacobian limit, therefore corresponding to decreasing missing masses. The open circles correspond to data integrated by two channels.

\begin{center}
\begin{table}[ht]
\begin{tabular}{c c c c c c c c c c c c c}
\hline
SPES3 mass&1004&1044&1094&1136&1173&1249&1277&1339&&1384&1479\\
marker&($\alpha$)&($\beta$)&($\gamma$)&($\delta$)&($\epsilon$)&($\eta$)&
($\phi$)&($\nu$)&&&\\
SPES4 mass 0.8$^{\circ}$&&1052&1113&1142&1202&1234&1259&&1370&1394&1478\\
SPES4 mass 2$^{\circ}$&996&1036&1104&1144&1198&1234&&1313&1370&&1477\\
\hline
\hline
SPES3 mass&1505&1517&1533&1542&(1554)&1564&1577&&&&\\
SPES4 mass 2$^{\circ}$&1507&1517&1530&1543&1557&1569&1580&&&&\\
\hline
\end{tabular}
\caption{Masses (in MeV) of narrow exotic baryons, observed previously in
SPES3 data and extracted from previous p($\alpha,\alpha$')X spectra measured at SPES4 \protect\cite{morsch1,morsch2}.}
\label{Table1}
\end{table}
\end{center}

A nice fit is obtained after introduction of the narrow resonances, leading to an important decrease of the $\chi^{2}$ (see Table 2).
\begin{center}
\begin{figure}[!ht]
\caption{(Color online.) Cross-section of the  pp$\to$n$\Delta^{++}$ reaction at $T_{p}$=0.79~GeV  \protect\cite{mosbacher,prout}. Inserts (a) and (b) correspond respectively to $\theta_{n}=0^{\circ}$ and $7.5^{\circ}$.}
\label{fig8}
\end{figure}
\end{center}
\subsection{The $^{2}$H(p,n)X reaction}
The differential cross-sections  $d^{2}\sigma/dE_{n}d\Omega_{n}$ of the $^{2}$H(p,n)X reaction were also measured at LAMPF \cite{prout}. The analysis was performed by the same authors \cite{mosbacher}, using the distorted-wave  impulse approximation. The result of the reanalysis is shown in Fig. \ref{fig9}.
Here the width of the structures is fixed at the width extracted from the pp$\to$n$\Delta^{++}$ data ($\sigma$=11~MeV), widen by the Fermi momentum taken to be p$_{F}$=66~MeV/c. Inserts (a), (b), and (c) show the result of the present reanalysis corresponding respectively to $\theta=0^{\circ}$, 7.5$^{\circ}$, and 15$^{\circ}$. The data are all integrated by two channels.  We observe that the present reanalysis allows to describe quite well the data; the total shapes
in the range between the $n$ and the $\Delta$ missing masses, are better reproduced after introducing the narrow resonances. The large improvement in $\chi^{2}$ is visible in Table 2. 

The comparison between pp$\to$n$\Delta^{++}$ and $^{2}$H(p,n)$\Delta^{++}$
reanalysis, allows us to favor isospin T=1/2 for the exotic narrow baryons at M=1004~MeV, 1044~MeV, and 1094~MeV, just as the baryonic structure at M=1277~MeV, since here the cross-section on D$_{2}$ target is much larger than on H target. The same cross-sections are extracted from both targets for the structures at M=1173~MeV and 1210~MeV, suggesting isospin T=3/2 for both of them. These deductions agree with the isospin values deduced previously \cite{bor2} with the help of a quark cluster mass formula which will be recalled  in Section X.
\begin{center}
\begin{figure}[!ht]
\caption{(Color online.) Cross-section of the  $^{2}$H(p,n)$\Delta^{++}$ reaction at $T_{p}$=0.79~GeV  \protect\cite{mosbacher,prout}. Inserts (a), (b), and (c) correspond respectively to $\theta_{n}=0^{\circ}$, 7.5$^{\circ}$, 
and 15$^{\circ}$.}
\label{fig9}
\end{figure}
\end{center}
\section{The  (d,2p) reactions} 
\subsection{The p(d,2p)$\Delta^{0}$ reaction.}
The cross-section of the p(d,2p)$\Delta^{0}$ reaction was measured with the SPES4 spectrometer at Saturne, using 2~GeV and 1.6~GeV incident deuterons, at the following angles: $\theta_{lab.}=0^{\circ}$, 0.5$^{\circ}$, 2.1$^{\circ}$, 4.3$^{\circ}$, and
7.2 $^{\circ}$ \cite{ellegaarde,jorgensen,sams}.
Inserts (a) and (b) of Fig. \ref{fig10} show the data respectively for
 $\theta$=0.5$^{\circ}$ and 4.3$^{\circ}$. They exhibit oscillating behaviour in the low part of the $\Delta$ missing mass range which are well fitted by the masses of the narrow baryonic structures. Inserts (a) and (b) of Fig. \ref{fig11} show the cross-sections of the same reaction, for respectively 
 $\theta$= 2.1$^{\circ}$ and 5.7$^{\circ}$. Here again, a nice fit for both angles, is obtained in nearly the whole range, when the narrow structures are added to the broad $\Delta$ peak.
 \begin{center}
\begin{figure}[!hb]
\caption{(Color online.) Cross-section of the  p(d,2p)$\Delta^{0}$  reaction at $T_{d}$=2~GeV $\theta=0.5^{\circ}$, and 4.3$^{\circ}$,  respectively in inserts (a), and (b), \protect\cite{ellegaarde,sams}.}
\end{figure}
\label{fig10}
\end{center}
\begin{center}
\begin{figure}[!ht]
\caption{(Color online.) Cross-section of the  p(d,2p)$\Delta^{0}$  reaction at $T_{d}$=2~GeV $\theta=2.1^{\circ}$, and 5.7$^{\circ}$, respectively in inserts (a), and (b) \protect\cite{ellegaarde,jorgensen,sams}.}
\label{fig11}
\end{figure}
\end{center}
\subsection{The $^{12}$C(d,2p)X reaction.}
Such reaction allows to observe cross-sections corresponding to quasi-free scattering on nucleons in the $^{12}$C target. The narrow peaks, corresponding to excited levels of $^{12}$B, are restricted to the energy loss
$\omega\le$50~MeV. This value ranges inside the large peak of the left part of Fig. \ref{fig12}. Above this mass, all peaks having total widths smaller than 40~MeV, are attributed to quasi-free production of narrow structures on nucleon targets. The total widths of the narrow baryonic structures are taken as $\Gamma_{t}\approx$20~MeV; they are slightly enlarged here, compared to previous reanalyses, due to the Fermi momentum of the target nucleon.

Two small peaks at M$\approx$1004~MeV and 1044~MeV in the quasi-free reaction, are introduced in order to improve the fit. They would correspond to E$_{exc.}$=60~MeV and 80~MeV, if attributed to $^{12}$B. In Fig. \ref{fig12}, the oscillations are less pronounced than those in Fig. \ref{fig11}.  At $T_{p}$=2~GeV $\theta=4.3^{\circ}$, - not illustrated - indications of small structures exist at masses close to  M$\approx$1004~MeV and 1044~MeV.
\begin{center}
\begin{figure}[!ht]
\caption{(Color online.) Spectra of the  $^{12}$C(d,2p)X  reaction at $T_{p}$=2~GeV $\theta=0^{\circ}$ \protect\cite{sams}. (No vertical scale in the original paper).}
\label{fig12}
\end{figure}
\end{center}
\section{The ($^{3}$He,t) reactions}
\subsection{The p($^{3}$He,t)$\Delta^{++}$ reaction.}
The cross-section of the  p($^{3}$He,t)$\Delta^{++}$ reaction 
was measured at many incident energies, and spectrometer angles, mostly at the SPES4 spectrometer at Saturne, but also at other places. The data from Dubna \cite{ableev}, at T$_{p}$=2.39~GeV, 4.53~GeV, and 8.33~GeV are not enough precise since their binning was taken as $\Delta$E=25~MeV. They are not reanalyzed here. The $^{3}$He(p,t)$\Delta^{++}$ reaction was studied at SPES1 (Saturne) at forward angle in the laboratory system, with the motivation to reduce the contribution of the direct graph and study the possible ($\Delta^{++}$,2n) component of the $^{3}$He wave function \cite{spes1}. The beam energy was T$_{p}$=850~MeV. These data were taken at 
$\theta_{lab.}=6^{\circ}$, 10$^{\circ}$, and 15$^{\circ}$; their cross-sections decrease fast with increasing angle, therefore only the cross-section of the smaller angle is shown here, in Fig. \ref{fig13}. 

We observe that indeed oscillations can be seen and are well fitted by the narrow structures (see Table 2) if the data are shifted by $\Delta$M=-9~MeV. This systematic shift may correspond to a possible saturation of the spectrometer magnetic field  B, as small as $\delta$p/p=6 10$^{-3}$, possibly not taken into account for large triton momenta for the SPES1 spectrometer: p$_{t}\approx$1.85~GeV/c. The shift may be due also to a bad knowledge of the incident beam energy, since for M$_{X}$=1250~MeV, $\partial M_{X}/\partial T_{p}$=1.24 or to a bad knowledge of the spectrometer angle, since here $\partial  M_{X}/\partial \theta_{t}$=-223.6 MeV/radian.

\begin{table}[ht]
\begin{tabular}{c c c c c}
\hline
Fig.&reaction&$\chi^{2}_{D}$&$\chi^{2}_{NR}$&Comment\\
\hline
7&pp$\to$n$\Delta^{++}$&4.7&2.1& \\
8(a)&pp$\to$n$\Delta^{++}$&3.6&0.9& \\
8(b)&pp$\to$n$\Delta^{++}$&6.8&1.3& \\
9(a)&d(p,n)X&44.8&2.3&for M$_{X}\ge$1~GeV\\
9(b)&d(p,n)X&58.7&1.6&for M$_{X}\ge$1~GeV\\
9(c)&d(p,n)X&20.6&0.8&for M$_{X}\ge$1~GeV\\
13&$^{3}$He(p,t)$\Delta^{++}$&2.5&1.5&\\
17(a)&$d(^{3}$He,t)X&12.1&3.0&\\
17(b)&$d(^{3}$He,t)X&17.2&1.1&\\
18(a)&$d(^{3}$He,t)X&14.7&2.2&\\
18(b)&$d(^{3}$He,t)X&53.0&4.0&\\
\hline
\end{tabular}
\caption{Selection of several statistical significances between data and broad PDG $\Delta$ resonance ($\chi^2_{D}$) on the one hand, and between data and total spectra including the narrow resonances   ($\chi^2_{NR}$) on the other hand.}
\label{Table2}
\end{table}
Fig. \ref{fig14} shows the cross-sections of the p($^{3}$He,t)$\Delta^{++}$ reaction \cite{ellegaard} measured at Saturne (SPES4 beam line) at T$_{^{3}He}$=1.5~GeV, $\theta=3.5^{\circ}$ in insert (a), T$_{^{3}He}$=2~GeV, $\theta=5^{\circ}$ in insert (b), and T$_{^{3}He}$=2~GeV, $\theta=0^{\circ}$ in insert (c). 

The data are well fitted using the masses of narrow baryonic structures. Special 
attention has to be brought to the low mass spectra. Below M$_{X}$=1100~MeV, the data in Fig. \ref{fig14} (empty squares) are rescaled by a factor of five. We observe here negative and positive numberq resulting from the subtraction of quite poor statistics data obtained with a CH$_{2}$ and a C targets. 
\begin{center}
\begin{figure}[!ht]
\caption{(Color online.) Cross-section of the $^{3}$He(p,t)$\Delta^{++}$ reaction, at T$_{p}$=850~MeV, $\theta_{lab.}$=6$^{\circ}$ \protect\cite{spes1} after a $\Delta$M=-9~MeV shift (see text).}
\label{fig13}
\end{figure}
\end{center}
\begin{center}
\begin{figure}[!b]
\caption{(Color online.) Cross-section of the  p($^{3}$He,t)$\Delta^{++}$  reaction at
 $T_{p}$=1.5~GeV $\theta=3.5^{\circ}$ insert(a), 
 $T_{p}$=2~GeV $\theta=5^{\circ}$ insert (b), and 
 $T_{p}$=2~GeV $\theta=0^{\circ}$ insert (c) \protect\cite{ellegaard}.}
 \label{fig14}
\end{figure}
\end{center}
The narrow structures in this mass range were predicted to be preferably isoscalar \cite{bor1,bor2}, therefore they cannot be excited in the reaction on proton target. The proton missing mass spectra, excited from the neutrons of $^{12}$C
and CH$_{2}$ targets, are obtained with large statistics, allowing the renormalization and the subtraction.
The smaller statistics obtained above the proton missing mass range may explain the observed behaviour. Indeed in inserts (b) and (c) the small bumps agree with the $\alpha$ and $\beta$ arrows.
\begin{center}
\begin{figure}[!ht]
\caption{Cross-section of the p($^{3}$He,t)X reaction at  T$_{^{3}He}$=2~GeV,  $\theta=0.25^{\circ}$ in insert (a) and $\theta$= $1.6^{\circ}$ in insert (b) \protect\cite{beatrice}. The small empty circles (blue) describe the theoretical calculation (see text) \protect\cite{beatrice}. Above this calculation the baryonic structures allow to fit the experimental data. The total spectrum is also shown (red curve).}
\label{fig15}
\end{figure}
\end{center}
Another set of p($^{3}$He,t)X reaction cross section data was also obtained at SPES4 \cite{beatrice} with 
T$_{^{3}He}$=2~GeV incident beam at the following laboratory angles:  $\theta=0.25^{\circ}$, 1.6$^{\circ}$, 2.7$^{\circ}$, and 4$^{\circ}$. The data are read and reanalyzed in Fig. \ref{fig15} and \ref{fig16}. The  calculation was performed by the authors \cite{beatrice}, including the A=3 form factors. The result of this calculation is taken as "background", namely we use the narrow structures to fit the difference between the  and the calculation. Fig. \ref{fig15} shows the results at $\theta=0.25^{\circ}$ in insert (a) and  $\theta=1.6^{\circ}$ in insert (b). 
Fig. \ref{fig16} shows the corresponding results for  $\theta=2.7^{\circ}$ in insert (a) and  $\theta=4^{\circ}$ in insert (b).

The fast decrease of the $\Delta^{++}$(1220~MeV) excitation, with increasing angles, allows us to extract narrow baryonic structures. The accuracy between the narrow structure masses extracted here, and those extracted previously, is noteworthy. However in the present case the strutures are broader.
\begin{center}
\begin{figure}[!ht]
\caption{ (Color online) Cross-section of the p($^{3}$He,t)X reaction at  T$_{^{3}He}$=2~GeV,  $\theta=2.7^{\circ}$ in insert (a) and $\theta$= 4$^{\circ}$ in insert (b) \protect\cite{beatrice}. The small empty circles (blue) describe the theoretical calculation (see text) \protect\cite{beatrice}.  Above this calculation the baryonic structures allow to fit the experimental data. The total spectrum (red curve) is also shown.}
\label{fig16}
\end{figure}
\end{center}
\begin{center}
\begin{figure}[!ht]
\caption{(Color online). Cross-section of the d($^{3}$He,t)X reaction at  T$_{^{3}He}$=2~GeV,  $\theta=0.25^{\circ}$ in insert (a) and $\theta$= 1.6$^{\circ}$ in insert (b) \protect\cite{beatrice}. The small empty circles describe the theoretical calculation (see text) \protect\cite{beatrice}. Above this calculation the baryonic structures allow to fit the experimental data.}
\label{fig17}
\end{figure}
\end{center}
\subsection{The d($^{3}$He,t)X reaction}
The d($^{3}$He,t)X reaction was studied at SPES4 (Saturne) at T$_{^{3}He}$=2~GeV and four different angles: $\theta=0.25^{\circ}$, 1.6$^{\circ}$, 2.7$^{\circ}$, and 4$^{\circ}$ \cite{beatrice}. The data are read and reanalyzed in Figs. \ref{fig17} and \ref{fig18}. The complete calculation was performed by the authors \cite{beatrice}, including the quasi-elastic contribution and the final state interaction ($\Delta$-N interaction and $\Delta$ N $\to$ NN process). The result of this calculation is taken as background, namely we fit the difference between data and the calculation with the function summing the gaussians corresponding to the narrow structures. Fig. \ref{fig17} shows the results at  $\theta=0.25^{\circ}$ in insert (a) and  $\theta=1.6^{\circ}$ in insert (b).
Fig. \ref{fig18} shows the corresponding results for  $\theta=2.7^{\circ}$ in insert (a) and  $\theta=4^{\circ}$ in insert (b). Whereas the H($^{3}$He,t)X reaction allows to excite only isospin I=3/2 final states, the  d($^{3}$He,t)X reaction allows to excite just as well I=3/2 as I=1/2 final states. 
The comparison between both reactions shows that, although the reaction on a proton target is favoured versus the reaction on a neutron target by a factor of three due to isospin couplings, the narrow baryonic structures are better excited on the deuterium target. We deduce as above that isospin I=1/2 is favoured for the low mass baryonic structures.

The global fits, after the introduction of the narrow structures in Fig. \ref{fig17}(a), \ref{fig17}(b), \ref{fig18}(a), and \ref{fig18}(b), are very good. Table 2 shows the related quantitative information.  
\begin{center}
\begin{figure}[!ht]
\caption{(Color online.) Same caption as in Fig. \protect\ref{fig17}, for  $\theta=2.7^{\circ}$ in insert (a) and $\theta= 4^{\circ}$ in insert (b) \protect\cite{beatrice}.}
\label{fig18}
\end{figure}
\end{center}
\begin{center}
\begin{figure}[!ht]
\caption{(Color online) Angular distributions of some narrow baryonic structures extracted from the  d($^{3}$He,t)X reaction. M=1044~MeV, full circle, black; M=1094~MeV, full square, red;
M=1136~MeV, full triangle, green; M=1213~MeV,  full star, blue; M=1247~MeV, empty circle, sky blue; M=1274~MeV, empty square, purple.}
\label{fig19}
\end{figure}
\end{center}
Fig. \ref{fig19} shows the angular distributions at forward angles of six narrow structures: M=1044~MeV,  
1094~MeV, 1136~MeV, 1213~MeV, 1247~MeV, and 1274~MeV. The error bars are set arbitrarily to 10\%. The cross-sections decrease continuously with increasing angle, more or less parallel to each other. These regular (non oscillating) shapes of the angular distributions, can be considered as an useful argument with respect to the present cross-section reanalysis.
\subsection{The $^{12}C(^{3}$He,t) reaction}
The $^{12}C(^{3}$He,t) reaction was studied at Saturne (SPES4 beam line)
\cite{contardo,oster,hennino} at T$_{^{3}He}$=2~GeV, $\theta=0^{\circ}$.
Fig. \ref{fig20} shows the result after adding the narrow structures to the theoretical analysis performed by \cite{oster}.
\begin{center}
\begin{figure}[!ht]
\caption{(Color online.) Cross-section of the  $^{12}C(^{3}$He,t) reaction at T$_{^{3}He}$=2~GeV, $\theta=0^{\circ}$. \protect\cite{contardo} 
\protect\cite{oster}. The text explains the content of the theoretical calculations performed by \protect\cite{oster}. Insert (a) shows the double differential cross-section; insert (b) shows the pion coincidence spectrum.}
\label{fig20}
\end{figure}
\end{center}

\vspace*{-0.8cm}
Insert (a) shows the double differential cross-section. The curve passing through the small empty circles is the final result of the calculation \cite{oster} within the isobar-hole model. It is obtained including the quasi-free $\Delta$ decay, the $\Delta$ spreading and the  coherent pion production. Insert (b) shows the pion coincidence spectrum 
\cite{hennino,thierry}. The theoretical curves show the result of the calculations \cite{oster} performed with Landau-Migdal parameters:
g$_{\Delta\Delta}$ = 0.33 (solid curve) and g$_{\Delta\Delta}$ = 0.4
 (dashed curve). This last curve was renormalized by a factor of 1.5 by the
 authors. The
narrow structures are extracted from the dashed curve. The extraction from the solid curve will change the amplitudes of these structures, but not their position. Here also we find the final fits show a remarkable agreement with the data.
\section{Charge exchange reactions with incident heavy ion beams}
\subsection{The p($^{12}$C,$^{12}$N)$\Delta^{0}$ reaction}
The p($^{12}$C,$^{12}$N)$\Delta^{0}$ reaction was studied at SPES4 (Saturne) using a $^{12}$C
beam of 1100~MeV/N at the spectrometer angle $\theta=0^{\circ}$ \cite{michele}.   The data are read, integrated over three channels and shown as a function of M$_{X}$.  Fig. \ref{fig21} shows these data, fitted with the $\Delta^{0}$ peak and the structures previously observed.
The important oscillatory behaviour is very well reproduced by the fit.   

\begin{center}
\begin{figure}[!ht]
\caption{(Color online.) Cross-section of the  p($^{12}$C,$^{12}$N)$\Delta^{0}$ reaction at $T_{p}$=1100~MeV/N $\theta=0^{\circ}$ \protect\cite{michele}.}
\label{fig21}
\end{figure}
\end{center}
\subsection{The p($^{20}$Ne,$^{20}$F)$\Delta^{++}$ and the 
 p($^{20}$Ne,$^{20}$Na)$\Delta^{0}$ reactions}
The p($^{20}$Ne,$^{20}$F)$\Delta^{++}$ and the 
 p($^{20}$Ne,$^{20}$Na)$\Delta^{0}$ reaction were measured at SPES4 (Saturne) with the  $^{20}$Ne beam at 900~MeV/N $\theta=0^{\circ}$  \cite{michele,gaarde}.  The data are read, and shown versus M$_{X}$ in  Fig. \ref{fig22}. The 
p($^{20}$Ne,$^{20}$F)$\Delta^{++}$ cross-sections, (insert (a)), are integrated over two  channels. The same width (FWHM=120~MeV), the same mass, and a ratio of three, determined by the isospin Clebsh-Gordan coefficients, are attributed to $\Delta^{++}$ and  $\Delta^{0}$. Such ratio supposes that the difference between the form factors of the final nuclei $^{20}$F and $^{20}$Na, differ only by a small amount. We observe small contributions of the first narrow baryonic masses - compatible with zero - in insert (a).
This result can be explained again by the expected isospin for these masses \cite{bor1}. In the same way, the large cross-section around M$\approx$1200~MeV in insert (a), agrees with the previous expectation, namely isospin=3/2. 
\begin{center}       
\begin{figure}[!ht]
\caption{(Color online.) Cross-section of the 
p($^{20}$Ne,$^{20}F)\Delta^{++}$  reaction (insert (a)), and  p($^{20}$Ne,$^{20}$Na)$\Delta^{0}$ reaction (insert (b)) with T$_{^{20}Ne}$=18~GeV, $\theta$=0$^{\circ}$ \protect\cite{michele,michele1,gaarde}.}
\label{fig22}
\end{figure}
\end{center}
\subsection{The  $^{12}$C( $^{12}$C, $^{12}$B)X reaction}
The experiments were performed at SPES4 (Saturne) on $^{12}$C \cite{michele1} and 
$^{208}$Pb targets (not shown). The spectra were studied at $\theta$=0$^{\circ}$, the beam energies being T$_{beam}$=900*A~MeV \cite{michele}\cite{michele2}. The reanalysis of the cross-sections on   $^{12}$C
target are shown in Fig. \ref{fig23}.
\begin{center}
\begin{figure}[!ht]
\caption{(Color online.) Cross-section of the 
 $^{12}$C($^{12}$C,$^{12}$B)X reaction (insert (a)),
$^{12}$C($^{12}$C,$^{12}$N)X reaction (insert (b)), and
$^{12}$C($^{20}$Ne,$^{20}$F)X  reaction (insert (c)). All three reactions were studied at $\theta=0^{\circ}$, and T$_{beam}$=900*A MeV
 \protect\cite{michele,michele1,gaarde,michele2}.}
 \label{fig23}
\end{figure}
\end{center}
The three inserts (a), (b), and (c) correspond respectively to the three reactions: $^{12}$C($^{12}$C,$^{12}$B)X,
$^{12}$C($^{12}$C,$^{12}$N)X, and
$^{12}$C($^{20}$Ne,$^{20}$F)X. In all three cases, we observe the need to introduce the structures, in addition to the $\Delta$ peak, to describe the data which are well fitted in that case. 
We observe small oscillations in the three spectra, at the positions corresponding to the two first masses: M=1004~MeV ($\alpha$) and M=1044~MeV ($\beta$). 
\subsection{The  $^{12}$C($^{14}$N,$^{14}$C)X reaction}
The  $^{12}$C($^{14}$N,$^{14}$C)X reaction was studied at SPES4 (Saturne) with
$^{14}$N incident beam of 880~MeV/A, and $\theta$=0$^{\circ}$ \cite{bache}. The result of our reanalysis is shown in Fig. \ref{fig24}. Again, the experimental oscillations are well reproduced with the narrow structure masses.
\begin{center}
\begin{figure}[!ht]
\caption{(Color online.) Cross-section of the  $^{12}$C($^{14}$N, $^{14}$C)X reaction studied at Saturne (SPES4) at $\theta=0^{\circ}$, and T$_{beam}$=880*A MeV.
 \protect\cite{bache}.}
 \label{fig24}
\end{figure}
\end{center}
\subsection{The p($^{16}$O,$^{16}$N)X and $^{12}$C($^{16}$O,$^{16}$N)X reactions}
The $^{16}$O,$^{16}$N transfer reactions were studied at Saturne (SPES4) on CH$_{2}$ and $^{12}$C targets \cite{gaarde1}. The reanalysis of both spectra is shown in Fig. \ref{fig25}.
\begin{center}
\begin{figure}[!ht]
\caption{(Color online.) Cross-section of the p($^{16}$O,$^{16}$N)X  reaction (insert (a)), and $^{12}$C($^{16}$O,$^{16}$N)X reaction (insert (b)). Both 
 reactions were studied at $\theta=0^{\circ}$, and T$_{beam}$=900*A MeV.
 \protect\cite{gaarde1}.}
 \label{fig25}
\end{figure}
\end{center}    
In the p($^{16}$O,$^{16}$N)X spectra, insert (a), the positions of the central peaks do not correspond so well to the expected masses as the in previous spectra. On the other hand,
the agreement in insert (b), describing the $^{12}$C($^{16}$O,$^{16}$N)X reaction
is outstanding, since
all peaks, including the structures at M=1004~MeV ($\alpha$) and M=1044~MeV
($\beta$) are perfectly reproduced. Here an arbitrary background is introduced in order to simulate the inelastic scattering due to the excitation of $^{12}$C levels.
\subsection{The  $^{27}$Al( $^{20}$Ne, $^{20}$Na)X and the  $^{27}$Al( $^{20}$Ne, $^{20}$F)X reactions}
The  $^{27}$Al($^{20}$Ne,$^{20}$Na)X and the  $^{27}$Al($^{20}$Ne,$^{20}$F)X reactions were measured at Saturne (SPES4), using a $^{20}$Ne incident beam of T=950~MeV/A, at $\theta=0^{\circ}$ \cite{bache}. The same analysis is done as above, based on the integration over two channels and the addition of narrow structures over a broad $\Delta^{++}$ peak. The result is shown in Fig. \ref{fig26}. Here, again, the oscillations are well reproduced, especially in the region between the nucleon and the $\Delta$ missing masses in insert (a).
\begin{center}
\begin{figure}[!ht]
\caption{(Color online.) Cross-section of the 
 $^{27}$Al( $^{20}$Ne, $^{20}$Na)X reaction (insert (a)), and
$^{27}$Al( $^{20}$Ne, $^{20}$F)X reaction (insert (b)). Both 
 reactions were studied at $\theta=0^{\circ}$, and T$_{beam}$=950*A MeV.
 \protect\cite{bache}.}
 \label{fig26}
\end{figure}
\end{center}
\section{Conclusion}
This work analyses different cross-sections measured with incident hadrons, in order to increase the experimental signatures for the excitation of low mass narrow baryons previously observed. A large improvement is obtained in the description of the data by a fit, which takes narrow structures into account. The corresponding $\chi^{2}$ values, which show such improvement, have been reported in Table 2. Most of the data were shown without preselecting those which present the most pronounced oscillations, well fitted by the exotic baryonic structure masses. 

This paper does not pretend to give an exhaustive outline of all existing data.

These structures were associated with quark clusters \cite{bor1} \cite{bor2}. This can explain the fact that reactions performed with incident leptons, which in principle allow similar observations  \cite{cepj}, lead to smaller signals than reactions performed with incident hadrons. In this case the number of quarks involved in the reactions are  reduced.

 All reactions reanalyzed here were measured in view of another physical goal, therefore the binning and the statistics were not optimized for the present study. So, there is here no attempt to study the mass range 950$\le$M$\le$1000~MeV, although several spectra
exhibit an excess of counting over the fits.  Moreover, this missing mass region, when composite targets are concerned, corresponds also to possible excitation of nuclei excited states.

In almost all spectra, clear structures are however observed, mainly in the mass region 1000$\le$M$\le$1130~MeV before the region where the $\Delta$ starts to dominate the countings. This mass range concerns the structures at
M=1004 and 1044~MeV, particularly interesting since they are both located at masses lower than the threshold of possible pion-nucleon disintegration. The narrow baryonic resonances are observed in almost all missing mass range studied. 
The most noteworthy spectra, are those showing notable oscillations well fitted by the narrow structures. This concerns a large part of the reactions discussed previously, namely the spectra corresponding to the pp$\to$p$\pi^{+}$X, (d,2p)X, (p,t)X, d($^{3}$He,t)X, p($^{12}$C,$^{12}$N)X,   
$^{12}$C($^{16}$O,$^{16}$N)X,
($^{20}$Ne,$^{20}$F)X, and ($^{20}$Ne,$^{20}$Na)X reactions.

The comparison between final states with isospin T=1/2 and T=3/2 leads us to favor isospin T=1/2 for the lower narrow structure masses. These structures are more easily observed in reactions using isoscalar scattered particles such as $\alpha$ or deutons.  We notice however that the overall results suggest possible isospin degeneracy.

In order to reproduce the masses of exotic, narrow, and weakly excited baryons, we use a phenomenological relation wich  describe them  by a configuration adding one (or more) $q{\bar q}$ pairs to the classical $qqq$ configuration. Such approach was justified theoretically \cite{diakonov2} in a paper which discussed the existence of the pentaquark: 
{\it In fact the Skyrme model is in accordance with quarks but it is in one way better as it allows baryons to be made of $N_{c}$ quarks, plus an indefinite number of $Q{\bar Q}$ pairs}.     

 Although theoretical studies must be carried out to describe these observations, it is useful to remind the nice mass spectra obtained with help of the phenomenological mass relation \cite{mulders} obtained within the assumption of two clusters of quarks at the ends of a stretched bag in terms of color magnetic interactions:
\begin{equation} 
M=M_0+M_1[i_1(i_1+1)+i_2(i_2+1)+(1/3)s_1(s_1+1)+                    
\end{equation}
\vspace*{-1.1cm}
\begin{equation} 
\hspace*{2.cm}(1/3)s_2(s_2+1)]
\end{equation}
Here $i$ and $s$ are isospins and spins of the two clusters. Notice that the formula involves degeneracy. We adjust the two parameters
$M_{0}$=838.2~MeV and  $M_{1}$=100.3~MeV in order to reproduce the masses, spins and isospins of the nucleon and of the Roper resonnace: $N^{*}$(1440). Then we  calculate the  masses of the other structures without any free parameter. These masses, possible spins and isospins, quark configurations, and parities if only s wave are considered, are shown in Fig. \ref{fig27}. Notice that the parity of the Roper N$^{*}$(1440) is positive if a P wave between $(qqq)$ and $q\bar q)$ quark clusters is assumed.

\begin{center}
\begin{figure}[ht]
\vspace*{2.cm}
\vspace*{8.cm}
\caption{(Color online.) Calculated properties of narrow exotic baryonic structures, using the phenomenological mass formula \protect\cite{mulders}: parity for s wave clusters, configurations as indicated, masses, possible spins and isospins. The right column shows the  experimetal masses.}
\label{fig27}
\end{figure}
\end{center}
Note also the attribution of isospin I=1/2 to the three lightest structure masses.

We conclude that the growing number of coherent indications reinforces the 
evidence for the genuine existence of these structures. These structures are clearly exotic since there is no room for them in the $qqq$ configurations \cite{caps}, since their width is smaller than the widths of "classical" PDG \cite{pdg} baryonic resonances, and also since some masses lie below the pion threshold mass. They are naturally associated with precursor quark deconfinement and may show that the quark and gluon degrees of freedom are effective even at very low energy and must be taken into account.

 

\end{document}